\documentclass[usenatbib]{mn2e}
\usepackage[utf8]{inputenc}
\usepackage{natbib,epsfig,cite,mystyle,amsmath,amssymb,graphicx}
\usepackage{array, xcolor, caption}
\usepackage[colorlinks=true,linkcolor=blue,citecolor=blue]{hyperref}

\usepackage[pass,letterpaper]{geometry}

\pdfoutput=1  % togliere il commento se le figure sono pdf, lasciarlo se sono eps
\usepackage{bm}
\usepackage{graphicx,color}
\usepackage{latexsym,amsmath,amssymb,graphicx,booktabs}
\usepackage{hyperref}
\usepackage{placeins}
\usepackage{subfigure}
\usepackage{mathtools}
\usepackage[normalem]{ulem}

\definecolor{MyBlue}{rgb}{0.15,0.15,0.70}
\definecolor{Dgreen}{rgb}{0,0.7,0.0}

\hypersetup{
colorlinks=true,
citecolor=MyBlue,
linkcolor=MyBlue,
urlcolor=MyBlue
}

\usepackage{amssymb}
\usepackage{amsmath}
\usepackage{amsfonts}
\usepackage{upgreek}
\usepackage{latexsym}
\usepackage{appendix}
\usepackage{eufrak}
\usepackage{dsfont}

\newcommand\ees{\end{eqnarray}}
\newcommand\bees{\begin{eqnarray}}

\usepackage[export]{adjustbox}

\newcommand\spart{\;\raise1.0pt\hbox{/}\hskip-6pt\partial}
\newcommand\spartb{\;\overline{\raise1.0pt\hbox{/}\hskip-6pt\partial}}

\newcommand{\be}{\begin{equation}}
\newcommand{\ee}{\end{equation}}

\newcommand{\beqa}{\begin{eqnarray}}
\newcommand{\eeqa}{\end{eqnarray}}

\newcommand{\m}{\rm{m}}

\newcommand{\dd}{\text{d}}

\newcommand{\Gal}{_{\rm G}}

\begin{document}
\title[SGWB anisotropies: astrophysical dependencies]{Stochastic gravitational wave background anisotropies in the mHz band: astrophysical dependencies}
\author[G. Cusin, I. Dvorkin, C. Pitrou, J.-Ph. Uzan]{Giulia Cusin$^{1}$\thanks{giulia.cusin@physics.ox.ac.uk}, Irina Dvorkin$^{2,3}$, Cyril Pitrou$^{3}$, Jean-Philippe Uzan$^{3}$ \\
$^{1}$Astrophysics Department, University of Oxford, DWB, Keble Road, Oxford OX1 3RH, UK \\
$^{2}$Max Planck Institute for Gravitational Physics (Albert Einstein Institute), Am M\"{u}hlenberg 1, Potsdam-Golm, 14476, Germany \\
$^{3}$Institut d'Astrophysique de Paris, Sorbonne Universit\'{e}, CNRS UMR 7095, 98 bis, Bd Arago, 75014 Paris, France \\
}

\pagerange{\pageref{firstpage}--\pageref{lastpage}} \pubyear{2019}
\maketitle
\label{firstpage}

\begin{abstract}
We show that the anisotropies of the astrophysical stochastic gravitational wave background in the mHz band have a strong dependence on the modelling of galactic and sub-galactic physics. 
%Anisotropies in the LISA band are studied here for the first time.  
We explore a wide range of self-consistent astrophysical models for stellar evolution and for the distribution of orbital parameters, all calibrated such that they predict the same number of resolved mergers to fit the number of detections during LIGO/Virgo O1+O2 observations runs. We show that different physical choices for the process of black hole collapse and cut-off in the black hole mass distribution give fractional differences in the angular power spectrum of anisotropies   up to 50\% on all angular scales. 
%Varying the distribution of initial semi-major axis at forma- tion induces relative differences on the angular power spectrum ? 70% on large angular separations. 
We also point out that the astrophysical information which can be extracted from anisotropies is complementary to the isotropic background and individual mergers. 
%We explore a wide range of astrophysical models for stellar evolution and black hole formation and show that they lead to fractional differences in the angular power spectrum of anisotropies (normalized over the monopole) of up to $\sim40\%$ on all angular scales. Varying the distribution of initial semi-major axis at binary formation leads to relative differences of the angular power spectrum of up to $\sim80\%$ at large angular separations. We also point out that the astrophysical information which can be extracted from anisotropies is complementary to the isotropic background and individual mergers.  
These results underline the interest in the anisotropies of the stochastic gravitational wave background as a new and potentially rich field of research, at the cross-road between astrophysics and cosmology.
\end{abstract}


%%%%%%%%%%%%%%%%%%%%%%%%%%%%%%%%%%%%%%%%%%%%%%%%%%
\section{Introduction} 
%%%%%%%%%%%%%%%%%%%%%%%%%%%%%%%%%%%%%%%%%%%%%%%%%%

The stochastic gravitational-wave (GW) background is generated by the superposition of various unresolved astrophysical and cosmological sources \citep{Allen:1996gp, Cornish:2001hg, Mitra:2007mc, Thrane:2009fp, Romano:2015uma, Romano:2016dpx, Renzini:2018vkx,TheLIGOScientific:2016xzw,2018arXiv181108797C}. Based on the recent observations of merging black holes (BHs) and neutron star binaries by the Advanced LIGO and Advanced Virgo detectors \citep{2015CQGra..32b4001A,2015CQGra..32g4001L,Abbott:2016blz,Abbott:2016nmj,TheLIGOScientific:2016pea,Abbott:2017gyy,Abbott:2017vtc,Abbott:2017oio,TheLIGOScientific:2017qsa,2018arXiv181112907T}, it is estimated that the stochastic background from unresolved stellar-mass binaries may be detected within a few years of operation of the LIGO-Virgo network \citep{Abbott:2017xzg,2019arXiv190302886T}. 
%The angular average (i.e. monopole) of the AGWB signal is defined as the energy density in GW per logarithmic frequency interval in units of the critical density of the Universe: $\Omega_{\rm GW}=\dd\rho_{\rm GW}/\dd\ln f/\rho_c$.
 Its anisotropic component is constrained by LIGO observations up to $\ell=4$ \citep{2019arXiv190308844T} which result in upper limits on the amplitude of the dimensionless energy density per units of logarithmic frequency in the range $\Omega_{\rm GW}(f=25 \text{Hz},\Theta)<0.64-2.47\times 10^{-8}$ sr$^{-1}$ for a population of merging binary compact objects, where $\Theta$ denotes the angular dependence. Methods to measure and map the astrophysical gravitational waves background (AGWB) in the LIGO and LISA (Laser Interferometer Space Antenna) frequency ranges are discussed in \citet{Allen:1996gp, Cornish:2001hg, Mitra:2007mc, Thrane:2009fp, Romano:2015uma, Romano:2016dpx, Renzini:2018vkx,TheLIGOScientific:2016xzw,2018arXiv181108797C}.

Since astrophysical sources reside in galaxies which cluster on large scales, it is expected that the AGWB will be direction-dependent, and thus include an anisotropic component. The anisotropies of the AGWB, as first derived in \citet{Cusin:2017fwz, Cusin:2017mjm}, and studied in \citet{Cusin:2018rsq, Cusin:2018avf, Cusinnew} and in \citet{Jenkins:2018uac,Jenkins:2018kxc}  \citep[see][for a critical analysis of these last two works]{Cusin:2018ump} depend on three key ingredients: {\it i)} the underlying cosmology {\it ii)} the large scale structure or galaxy clustering and {\it iii)} the local astrophysics on sub-galactic scales. The standard cosmological model and its parameters are now constrained with a high precision, and the evolution and properties of the large scale structure are also well understood at the scales relevant for the AGWB \citep[see][for details on this latter point]{Cusin:2018ump}. In this work we assume the standard cosmological model including structure formation using the values of the cosmological parameters from {\it Planck}~ \citep{Planck2018}.  The sub-galactic astrophysics, in particular the formation and evolution of binary compact objects, is however less well understood and more difficult to constrain with the usual electromagnetic observables.
% It The AGWB can potentially provide new information on the relevant astrophysical processes, such as stellar evolution, BH formation rate and mass distribution, as well as the properties of the host galaxies of GW sources. 

In this letter we show that the AGWB anisotropies are very sensitive to sub-galactic astrophysical modeling. In particular we show that different descriptions of stellar evolution and black hole binary formation lead to fractional differences in the angular power spectrum of anisotropies up to $\sim 50\%$, independently on the global normalization (monopole). 
We also identify a set of astrophysical functions (such as mass cut-off and initial mass function) which are not determining the shape of the angular power spectrum of anisotropies but only its total amplitude while other parameters (e.g. the initial distribution of the semi-major axis of the orbit) give an $\ell$-dependent rescaling of the power spectrum. 
Finally we demonstrate that monopole and anisotropies contain complementary astrophysical information and that studying the latter will allow one to break degeneracies between different astrophysical ingredients and potentially to constrain them separately.  
%This indicates the potential of studying AGWB anisotropies to set constrains on this sub galactic astrophysical functions.}
%different models of stellar evolution and collapse and on the distribution of initial separations of binary systems. }

\vspace{-2mm}
%%%%%%%%%%%%%%%%%%%%%%%%%%%%%%%%%%%%%%%%%%%%%%%%%%
\section{Astrophysical building blocks} 
%%%%%%%%%%%%%%%%%%%%%%%%%%%%%%%%%%%%%%%%%%%%%%%%%%

The quantity traditionally used to characterize the AGWB signal is the energy density in GW per logarithmic frequency interval in units of the critical density of the Universe, averaged over directions: $\Omega_{\rm GW}=\dd\rho_{\rm GW}/\dd\ln f/\rho_c$. It can also be written as the sum of contributions from sources located at all the (comoving) distances $r$ in the form $\Omega_{\rm GW}(f)\equiv\bar \Omega(f) = \int \partial_r \bar \Omega(f,r) \dd r$. In an homogeneous and isotropic background 
\be\label{link}
\partial_{r}\bar{\Omega}=\frac{f}{\rho_c} a^4 \int \dd\theta_{\Gal}\bar{n}_{\Gal}(r, \theta_{\Gal})\mathcal{L}_{\Gal}(r, f_{\Gal}, \theta_{\Gal})\,, 
\ee
where $\mathcal{L}_{\Gal}$ is the effective GW luminosity of a galaxy per unit of emitted frequency, $f_{\Gal}$, characterized by the set of parameters $\theta_{\Gal}$ (mass, metallicity...) and $n_{\Gal}$ stands for the number density of galaxies. Each astrophysical model predicts a functional dependence of $\partial_r \bar \Omega(f,r)$. 
%The relation between the effective luminosity and the emitted strain is established in Eqs.~(79) and~(80) of Ref.~\citet{Cusin:2017mjm}.
% In Eq.\,(\ref{AA}), $n_{\Gal}$ is the comoving number density of galaxies, $a$ is the scale factor of the Friedman-Lema\^{\i}tre spacetime normalized to 1 today. 
In addition, in the Limber approximation, the general expression of the angular power spectrum of the anisotropies of the AGWB reduces to \citep{Cusin:2017fwz,Cusin:2018ump}\footnote{\textcolor{black}{We stress that we use here the Limber approximation only to display the formulae of the angular power spectrum in a way that makes it apparent the main ingredients entering the result and their discussion. We use the complete formulae for the angular power spectrum, derived in \citet{Cusin:2017fwz}, see also \citet{Pitrou:2019rjz}, in our analysis.}}
\be\label{unique}
C_\ell(f) \simeq  \left(\ell+\tfrac{1}{2}\right)^{-1}\int \dd k P(k) \left|\partial_r \bar  \Omega (f,r)\right|^2\,,
\ee
where $\ell$ is the multipole in the spherical harmonic expansion, $P(k)$ is the galaxy power spectrum, and with the constraint $k r = \ell+1/2$. Hence, the multipoles are sensitive to the shape of $\partial_r \bar\Omega$, and in particular to the low-redshift value for low $\ell$ \citep[see][for details and derivations]{Cusin:2018ump,Cusinnew}. 
%
%The latest analysis \citep{2019arXiv190302886T} of the first and second LIGO observing runs lead to the upper bound on the isotropic background of $\Omega_{\rm GW}(f=25 \text{Hz}) < 4.8\times 10^{-8}$, assuming a population of compact binary sources, and  $\Omega_{\rm GW}(f=25 \text{Hz}) < 6\times 10^{-8}$ for a frequency-independent background. {The anisotropic component is constrained by LIGO observations up to $\ell=4$ \citep{2019arXiv190308844T} which result in upper limits on the amplitude in the range $\Omega_{\rm GW}(f=25 \text{Hz},\Theta)<0.64-2.47\times 10^{-8}$ sr$^{-1}$ for a population of merging binary compact objects, where $\Theta$ denotes the angular dependence. Methods to measure and map the AGWB in the LIGO and LISA (Laser Interferometer Space Antenna) frequency ranges are discussed in \citet{Allen:1996gp, Cornish:2001hg, Mitra:2007mc, Thrane:2009fp, Romano:2015uma, Romano:2016dpx, Renzini:2018vkx,TheLIGOScientific:2016xzw,2018arXiv181108797C}

This {\em letter} computes the angular power spectrum of the anisotropies in the AGWB from merging stellar-mass binary BHs using the astrophysical framework described in \citet{Dvorkin:2016okx,Dvorkin:2016wac,2018MNRAS.479..121D}, which we summarize below (see the companion article~\citet{Cusinnew} for extensive explanations). For each model, we compute the GW luminosity $\mathcal{L}_{\Gal}(z, f_{\Gal}, M_{\Gal})$ of a given galaxy as a function of halo mass $M_{\Gal}$ and redshift $z$ as follows. First, we calculate the SFR, $\psi(M_{\Gal},t)$, and the stellar-to-halo mass ratio using a modified version of the abundance-matching relations of \citet{2013ApJ...770...57B}. We use a Salpeter-like IMF~\citep{1955ApJ...121..161S} to describe the number of stars per unit total stellar mass formed, $\phi=\dd N/\dd M_*\dd M_{\rm tot,*}\propto M_*^{-p}$, where $M_*$ is the mass of the star at formation. Massive stars collapse to BHs, and we assume that the remnant mass depends only on the mass of the progenitor star, $M_*$, and on its metallicity, $Z$. This is encoded in a function $m=g_s(M_*,Z)$, that needs to be computed for each model. We adopt the observational relation of \citet{2016MNRAS.456.2140M} for the metallicity of the interstellar medium as a function of galaxy stellar mass and redshift. We also introduce a cut-off at high BH masses $M_{\rm co}$, that also depends and the astrophysical model.

 Under these assumptions, the instantaneous BH formation rate at cosmic time $t$ (or, equivalently, redshift $z$)  for a galaxy with halo mass $M_{\Gal}$ in units of events per unit BH mass $m$ is given by ${\cal R}_1(m,t)=\psi[M_{\Gal},t] \phi(M_*)\times \dd M_*/\dd m$ where $M_*(m)$ and $\dd M_*/\dd m$ are deduced from the relation $m=g_s(M_*,Z)$. Then, we assume that only a fraction $\beta$ of these BHs resides in binary systems that merge within the age of the Universe, so that the rate of formation of the latter is ${\cal R}_2(m,t)=\beta {\cal R}_1(m,t)$. The overall factor $\beta$ is used to normalize our model with respect to the number of events observed by LIGO/Virgo. Following \citet{2018MNRAS.479..121D}, the birth rate of binaries with component masses $(m,m' \leq m)$ is ${\cal R}_{\rm bin}(m, m')= {\cal R}_2(m){\cal R}_2(m')P(m,m')$, where $P(m, m')$ is a normalized distribution describing the association of mass $m$ and $m'$ in a BH binary system.

%Since BH binaries are expected to circularize rapidly due to GW radiation reaction before reaching the LIGO/Virgo band \citep[see e.g.][]{2018PhRvD..98h3028L}, 
Since binary eccentricity is expected to be small in the isolated field formation scenario considered here, we assume circular orbits in what follows. The distribution of the delay (or merger) time, $\tau_{\rm m}(m, m', a_{\rm f})$, between formation and mergers can be expressed through a distribution $f(a_{\rm f})$ of the semi-major axis at the formation of the binary and of the binary mass distribution.  Then the birth rate of BH binaries (per unit mass squared per unit time and per units of $a_{\rm f}$) is $\mathcal{R}_f[m,m', a_{\rm f}, t]={\cal R}_{\rm bin}(m, m')f(a_{\rm f})$ from which we deduce the merger rate at time $t$,  $\mathcal{R}_{\m}[m, m', a_{\rm f}, t]= \mathcal{R}_f[m, m', a_{\rm f}, t-\tau_{\rm m}(m, m', a_f)]$. Finally, the quantity $\partial_r \bar \Omega$ is given by the total contribution of merging binary BHs in the entire galaxy population, weighted by the halo mass function $\dd n/\dd M_{\Gal} (M_{\Gal},z)$ given in \citet{2008ApJ...688..709T}.

As reference astrophysical model we choose a Salpeter-like IMF with slope $p=2.35$. The BH formation model $m=g_s(M_*,Z)$ was taken to be the "delayed'" model in \citet{2012ApJ...749...91F} with a cutoff mass of $M_{\rm co}=45M_{\odot}$ \citep{2018arXiv181112940T}. We assume $P(m, m')=$~cst for the distribution of masses in the binary, and a distribution of the semi-major axis \emph{at formation} of $f(a_{\rm f})\propto a_{\rm  f}^{-1}$ between $a_{\rm f,min}=0.014$ AU and $a_{\rm f,max}=4000$ AU. The lower bound was chosen so as to ensure that the lightest BH binaries in our model [$(5,5) M_{\odot}$] merge within a Hubble time. The normalization $\beta$ is adjusted so as to match the number of detections by aLIGO/aVirgo during the O1+O2 observing runs \citep{2018arXiv181112907T}. We therefore require that all our models result in $10$ detectable events over the span of the O1+O2 observation time. Our estimate of the detection rates follows \citet{2018MNRAS.479..121D}, namely we calculate the signal-to-noise ratio (SNR) for each binary BH merger produced in the model using the noise power spectral densities from \citet{LIGOcurves1,LIGOcurves2} and a correction factor following \citet{1993PhRvD..47.2198F} to account for different source orientations. The GW strain is calculated using the PhenomB template \citep{2011PhRvL.106x1101A} assuming zero spins. We define observed events as those with SNR $>8$. The number of sources detectable during O1+O2 is given by multiplying the detection rate by the total observation time $T_{\rm obs}=169.7$ days.  Again, we emphasize that while individual mergers are resolved only at low $z$, the AGWB depends on the whole redshift distribution up to high redshifts. Therefore, even though all the models are calibrated to predict the same number of resolved sources, their resulting AGWB may vary if the high-redshift population of sources differs among the models.

\begin{figure}
\includegraphics[width=\columnwidth]{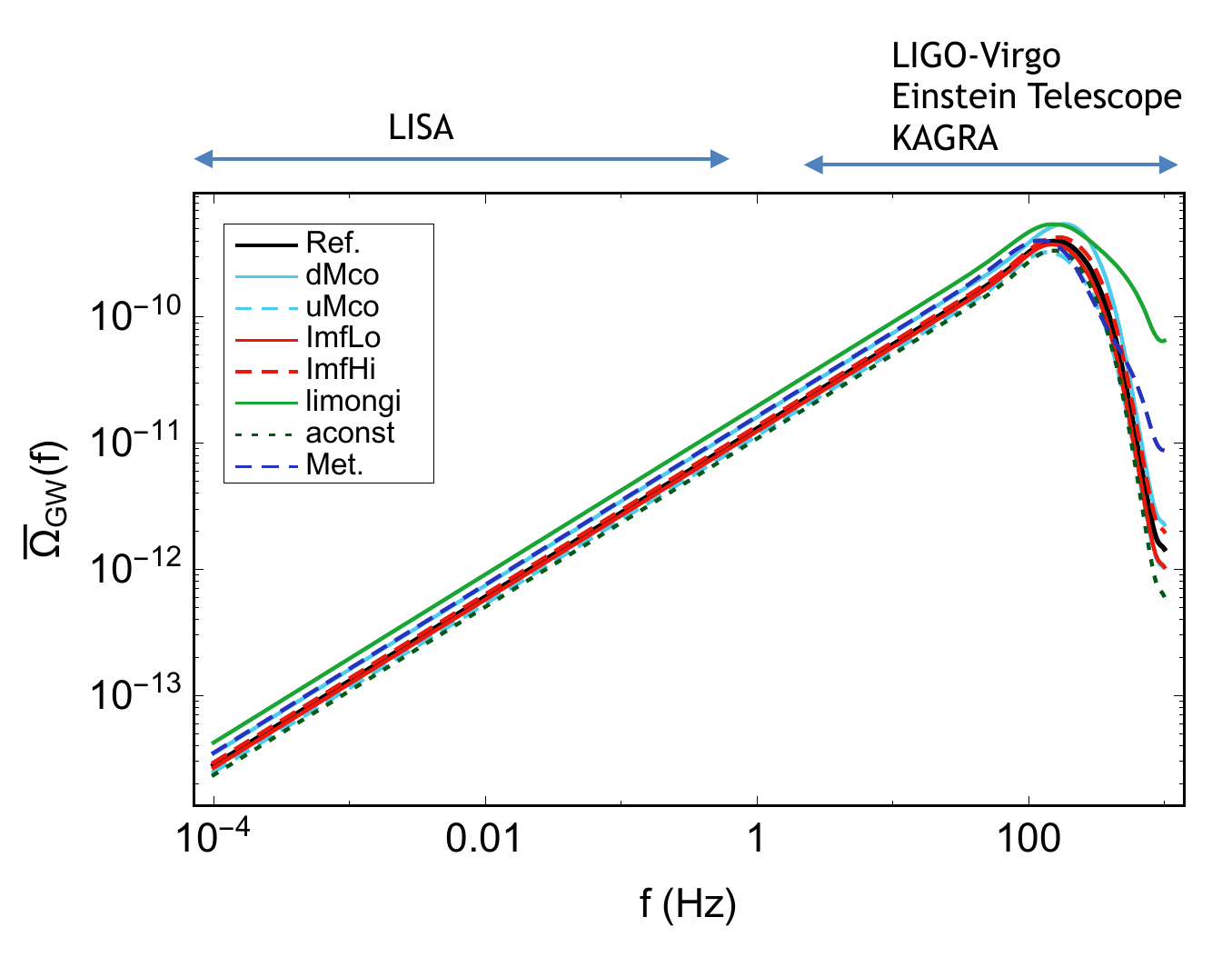}
\caption{Monopole of the AGWB for the different astrophysical models in this letter.  
[\emph{dMco}/\emph{uMlco} for the BH mass cut-off; \emph{imfHi}/\emph{imfLo} for the IMF slope; \emph{limongi} for the BH formation model; \emph{aconst} for $f(a_{\rm f})$; \emph{Met} for constant metallicity].}\label{Fig2}
\end{figure}

\begin{figure}
\includegraphics[width=\columnwidth]{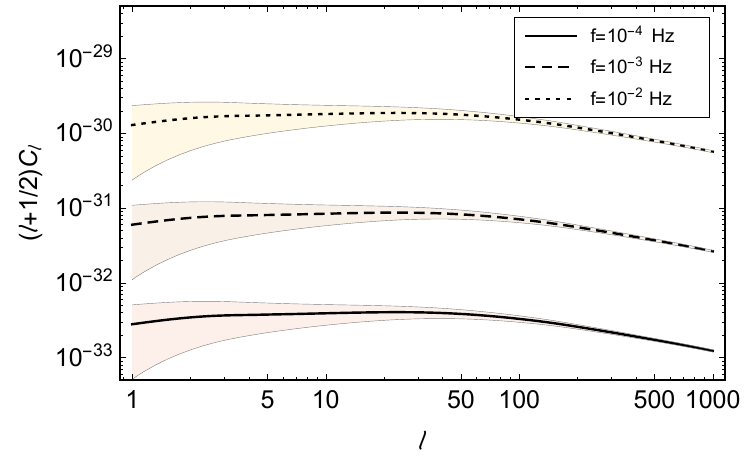}
\caption{Angular power spectrum of anisotropies for three frequencies in the mHz band. Multiplication by $\ell+1/2$ emphasizes the large-scale behaviour of Eq.\,\eqref{unique}. }\label{Fig4}
\end{figure}

\begin{figure*}
\includegraphics[width=\columnwidth]{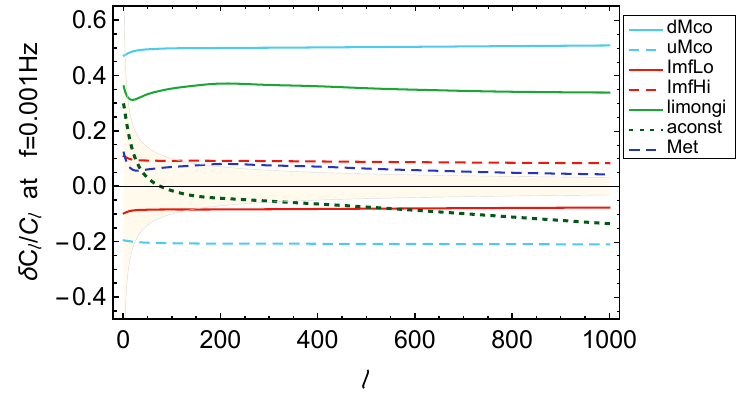}
\includegraphics[width=0.9\columnwidth]{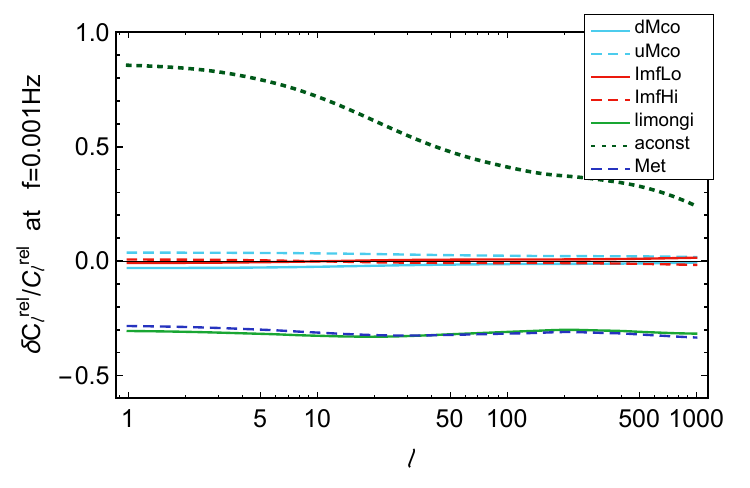}
\caption{Fractional difference between the angular power spectrum of anisotropies in different astrophysical models and in the reference model at $f=10^{-3}$~Hz (left). Fractional difference between angular power spectra, normalized over the monopole (right). }\label{Fig3}
\end{figure*}

\vspace{-2mm}
%%%%%%%%%%%%%%%%%%%%%%%%%%%%%%%%%%%%%%%%%%%%%%%%%%
\section{Astrophysical models} 
%%%%%%%%%%%%%%%%%%%%%%%%%%%%%%%%%%%%%%%%%%%%%%%%%%

We consider a series of models, each varying from the reference model described above in one key aspect.

\noindent{\bf BH formation model.} BH masses depend on the properties of their stellar progenitors (mostly their masses, chemical composition and rotation velocity e.g. \citet{2011ApJ...730...70O,2016MNRAS.458.2634M,2016A&A...588A..50M,2017hsn..book..513L}). The formation of \emph{binaries} is further influenced by common envelope evolution in the case of isolated binaries, and various dynamical processes in the case of stellar clusters~\citep{2012ApJ...759...52D,2016A&A...594A..97B,2016PhRvD..93h4029R}. In this letter, we assume that binary formation process is encoded in the efficiency parameter $\beta$ and the distribution of merger time delays. Furthermore, we restrict this study for simplicity to only one aspect of this complex problem, namely the evolution of isolated massive stars.  

Our reference model uses the description of \citet{2012ApJ...749...91F} which provides an analytic model for a neutrino-driven explosion and calculate the explosion energy, as well as the remnant mass, using numerical pre-collapse stellar models from \citet{2002RvMP...74.1015W}.  Another set of stellar evolution models is provided in \citet{2017hsn..book..513L}. These models differ from those in \citet{2012ApJ...749...91F} in two aspects. First, \citet{2017hsn..book..513L} uses a different set of pre-collapse stellar models which vary from \citet{2002RvMP...74.1015W}  in their treatment of convection, mass-loss rate and angular momentum transport. Second, \citet{2017hsn..book..513L} assumed a constant explosion energy in the calculation of the remnant mass, contrary to \citet{2012ApJ...749...91F}. As shown in \citet{2018MNRAS.479..121D}, these models predict different mass distributions of detectable BHs. In the following, the model \emph{limongi} uses the model described in \citet{2017hsn..book..513L} without stellar rotation.

\noindent{\bf BH mass cutoff.} Very massive stars (typically in the range $130-260M_{\odot}$) are unstable to electron-positron pair creation which may lead to a pair-instability supernova (PISN) that disrupts the entire star and prevents the formation of a BH \citep{2011ApJ...734..102K}. The absence of BHs in the mass range $60-260M_{\odot}$ may provide an indirect confirmation of this effect. A cutoff in BH mass may be present at even lower masses due to pulsational PISN, where the instability causes short episodes of mass ejection followed by periods of quiescent evolution \citep{2007Natur.450..390W,2015ASSL..412..199W}. As a result, the stellar mass is reduced below the limit of the onset of the instability. It was suggested in \citet{2018ApJ...856..173T} that this process may lead to an excess of BHs around $\sim 40M_{\odot}$. Recent analysis of the LIGO/Virgo events detected during the O1+O2 observational runs \citep{2018arXiv181112940T} provides a tentative measurement of the BH mass cutoff at $M_{\rm co}=45M_{\odot}$.  In order to explore the sensitivity of the stochastic background to the PISN-induced mass cutoff we varied it from $M_{\rm co}=40M_{\odot}$ [\emph{dMco}-models] to $M_{\rm co}=50M_{\odot}$ [\emph{uMco}-models].

\noindent{\bf Stellar initial mass function.} Our reference model assumes a Salpeter-like IMF with slope $p=2.35$. Interestingly, some studies suggest that the IMF slope may not be universal (see e.g. the discussion in \citet{2014PhR...539...49K}), for example a recent hint to a more shallow IMF in the Large Magellanic Cloud \citep{2018A&A...618A..73S}. In order to estimate the influence of the IMF we explore two models where the slope was taken to be $p=2.6$ [\emph{imfHi}-models] and $p=2.1$ [\emph{imfLo}-models].

\noindent{\bf Distribution of initial separations.} The reference model assumes the initial semi-major axis of the BH binaries to be distributed according to $f(a_{\rm f})\propto a_{\rm f}^{-1}$. This translates into a distribution of merger delay times of $f(t_{\rm delay})\propto t_{\rm delay}^{-1}$, favoring short delay times. We also consider an extreme scenario [\emph{aconst-model}] of flat distribution of the initial separations $f(a_{\rm f})\sim$~const, which results in longer delay times. 

\noindent{\bf Metallicity} Metal-rich stars experience strong winds which can considerably reduce their masses. Therefore, the masses of the remnant BHs are very sensitive to the metallicity of the progenitors stars \citep[e.g.][]{2012ApJ...749...91F}. This effect translates into a dependence on host galaxy mass and redshift, since low-mass and/or high-redshift galaxies typically contain less metals. To explore this dependence we considered a model with a constant metallicity of $10^{-3}Z_{\odot}$, which corresponds to an early stellar population. 

\vspace{-2mm}
%%%%%%%%%%%%%%%%%%%%%%%%%%%%%%%%%%%%%%%%%%%%%%%%%%
\section{Results} 
%%%%%%%%%%%%%%%%%%%%%%%%%%%%%%%%%%%%%%%%%%%%%%%%%%

Figure~\ref{Fig2} depicts the AGWB monopole as a function of frequency (in LISA and LIGO/Virgo frequency bands), for the different astrophysical models explored in this letter. Figure~\ref{Fig4} shows prediction of the angular power spectrum of anisotropies in the mHz band, for the reference astrophysical model. Since the contribution to this frequency band from merging stellar-mass BHs is from the inspiralling part of the waveform, the frequency dependence of the angular power spectrum is exactly the same as of the monopole so that  the relative anisotropies are frequency independent, see \citet{Cusinnew} for a detailed discussion of this latter point. We added the contribution of cosmic variance as a shaded region around each curve. This is the only  \emph{theoretical noise contribution} (in the sense that it does not depend on any instrument specification) present for a stationary and continuous background. \footnote{The contribution of space-like shot noise is expected to dominate the signal only at high multipoles for realistic choices of cut-off in flux, see \citet{Cusinnew} for a detailed discussion and estimate of this noise contribution.}  Fig.~\ref{Fig3} (left panel) presents the relative difference
\be
\delta C_{\ell}/C_{\ell}\equiv(C_{\ell}^{\rm mod}-C_{\ell}^{\rm ref})/C_{\ell}^{\rm ref}\,,
\ee
between the angular power spectra of each model (mod) and the reference model (ref) at the frequency of $f=10^{-3}$~Hz. With a shaded region around zero we represent the level of cosmic variance. It can be seen that the anisotropic component exhibits some differences, reaching 50\% even for the low multipoles. Note also that the shape of the spectra differs from the reference model for the modified $f(a_{\rm f})$ [\emph{aconst}] model, while the other models result in a renormalisation of the power independent of the multipole. Changing the BH cut-off mass $M_{\rm co}$ by $\sim 10\%$ results in a relative variation in anisotropies of up to 50\% for small multipoles ($\ell<10$), while changing the stellar evolution model generates a change of $\sim 40\%$ across the entire multipole range. This demonstrates that the astrophysical information contained in the monopole and in the spectrum of anisotropies is complementary. While the former is sensitive to the integral over redshift of the astrophysical kernel describing sub-galactic physics, anisotropies at different angular separations are sensitive to the amplitude of this kernel at different redshifts. This can be easily seen from Eq.~(\ref{unique}) where the Limber approximation has been used to relate distances to multipoles.  
%These results hold for both LIGO and LISA frequency bands. 
In order to disentangle the information contained in anisotropies from the global amplitude of the monopole, on the right panel of Fig.~\ref{Fig3} we plot fractional differences between angular power spectra normalized over the monopole (for each model). This plot shows that changing the distribution of the initial semi-major axis of the orbit leads to a $\ell$-dependent rescaling of the angular power spectrum. For the other models, differences in the astrophysical modeling lead to a \emph{global rescaling} of the angular power spectrum of anisotropies, different than the rescaling of the monopole.

%Different choices  of the distribution of semi axis and of the metallicity profile give  fractional differences $40\%$ while an even bigger role is played by the choice of the stellar evolution model. 

The fractional difference between angular spectra of different astrophysical models in the LIGO-Virgo frequency  band is very similar to that shown in Fig.~\ref{Fig3} for the mHz band, see e.g.  \citet{Cusinnew}. There is however an important caveat. The nature of the background in the LIGO-Virgo and LISA  bands is very different. At frequencies typical of LISA, the background is expected to be stationary and continuous over the time of observation  (as the typical duration of the inspiralling phase is much larger then the observation time), while the background in the LIGO-Virgo band is popcorn like. To properly model anisotropies in the LIGO band, one has to add a time-like shot-noise contribution to the signal part, which is flat in multipole space and adds as an offset to the pure signal part of the angular power spectrum, as discussed in \citet{Cusinnew} and also in \citet{Jenkins:2019uzp}. We stress that this noise component is absent in the mHz frequency band that we are studying in this work and that cosmic variance is the only source of theoretical noise at small angular separation.\footnote{In the LIGO band, for typical observation times of the order of few years, time-like shot noise dominates the signal. \textcolor{black}{Cross correlation with a galaxy map might be the most promising path to alleviate this shot noise problem affecting the LIGO-Virgo frequency band, as discussed for the first time in  \citet{Cusinnew}}.  In the LISA band where the background is continuous and stationary this time-like shot noise is absent and the spectrum presents only a space-like shot noise component due to the fact that galaxies have a discrete spatial distribution, which however affects the signal only at small angular separation, see \citet{Cusinnew}. This same shot-noise contribution is present  in the content of the CIB  see e.g. \citet{2011A&A...536A..18P}.}
%We also stress that cosmic variance has a cosmological origin. 
Since in all our predictions we are fixing a reference cosmology and we only vary the astrophysical modeling of sources, all the models in Fig.~\ref{Fig3}  are affected in the same way by cosmic variance.

%A full analysis is presented in the companion article~\citep{Cusinnew}. 

\vspace{-2mm}
%%%%%%%%%%%%%%%%%%%%%%%%%%%%%%%%%%%%%%%%%%%%%%%%%%
\section{Conclusions}
%%%%%%%%%%%%%%%%%%%%%%%%%%%%%%%%%%%%%%%%%%%%%%%%%%

This letter explored the astrophysical dependences of the anisotropies of the AGWB, considering  the contribution of stellar-mass binary BH mergers in both the LIGO/Virgo and LISA frequency bands.
%This letter clarifies the role of the astrophysical assumptions that underpin the various predictions of the angular power spectra.  
%We analyzed which ingredients of the astrophysical parametrization give sizable differences in the  anisotropies  and  which  do  not  play  a  role and we emphasized that 
Since cosmological structure formation processes are well-understood, all the modelling uncertainty is related to the formation and evolution of GW sources. 

We showed that the amplitude and shape of the angular power spectrum  are very sensitive to some astrophysical parameters such as the BH formation scenario, the cut-off in the mass distribution of the two black holes in the binary. In particular, this work is the first to identify the astrophysical functions (e.g. stellar initial mass function) that only give a global scaling to the spectrum 
of anisotropies, i.e. affect anisotropies and monopole in the same way. On the other hand, it points out that the role of metallicity, of the stellar evolution models and of the cut-off in the BH mass distribution produce a re-scaling of the spectrum non-degenerate with the scaling of the monopole. This result demonstrates that  the AGWB anisotropies are very sensitive to certain astrophysical parameters, contrary to previous claims presented in the literature. We have also discussed different sources of \emph{theoretical noise} (cosmic variance, shot-noise...) that need to be taken into account when modeling anisotropies and we have shown that variations due to astrophysical  modeling can be bigger than cosmic variance, which is the dominant source of theoretical uncertainty at mHz frequencies and for large angular separations. Hence in the mHz band there are no sources of theoretical noise that prevent one from accessing and measuring this new observable, with a sufficiently good sensitivity and angular resolution. This study is a first and necessary step towards a full and comprehensive analysis of the astrophysical content of this background. Further work should be done to understand the full power of AGWB and develop robust analysis techniques.

\vspace{-2mm}

%\FloatBarrier
\bibliographystyle{mn2e}

\bibliography{myrefs}

\label{lastpage}

\end{document}